\documentclass[aps,preprint,superscriptaddress,nofootinbib,tightenlines, 11pt]{revtex4-1}

\usepackage[utf8]{inputenc}
\usepackage[english]{babel}

\usepackage{graphicx}
\usepackage{epstopdf}

\usepackage{amsmath}
\usepackage{braket}
\usepackage{bbm}
\usepackage{xcolor}
\usepackage{color}
\usepackage{ulem}

\usepackage{hyperref}
\hypersetup{
    colorlinks,
    linkcolor={red!50!black},
    citecolor={blue!50!black},
    urlcolor={blue!80!black}
}
\newcommand{\Mhi}{M_{\text{hi}}}
\newcommand{\Mlo}{M_{\text{lo}}}

\begin{document}
\title{Halo structure of $^{17}$C}

\author{J. Braun}
\email{braun@theorie.ikp.physik.tu-darmstadt.de}
\affiliation{Institut für Kernphysik, Technische Universität Darmstadt,
64289 Darmstadt, Germany}

\author{H.-W. Hammer}
\email{Hans-Werner.Hammer@physik.tu-darmstadt.de}
\affiliation{Institut für Kernphysik, Technische Universität Darmstadt,
64289 Darmstadt, Germany}
\affiliation{ExtreMe Matter Institute EMMI, GSI Helmholtzzentrum
für Schwerionenforschung GmbH, 64291 Darmstadt, Germany}

\author{L. Platter}
\email{lplatter@utk.edu}
\affiliation{Department of Physics and Astronomy
University of Tennessee, Knoxville, TN 37996, USA}
\affiliation{Physics Division, Oak Ridge National Laboratory, Oak
  Ridge, TN 37831, USA}

\date{\today}

\begin{abstract}
  $^{17}$C has three states below the $^{16}$C~+~$n$ threshold
  with quantum numbers $J^P=3/2^+,1/2^+,5/2^+$.
  These states have relatively small neutron separation
  energies compared to the neutron separation and excitation energies of
  $^{16}$C. This separation of scales motivates our investigation of
  $^{17}$C in a Halo effective field theory (Halo EFT) with a $^{16}$C
  core and a valence neutron as degrees of freedom. We discuss
  various properties of the three states such as electric
  radii, magnetic moments, electromagnetic transition rates and
  capture cross sections. In particular, we give predictions for the
  charge radius and the magnetic moment of the $1/2^+$ state and for
  neutron capture on $^{16}$C into this state. Furthermore, we discuss
  the predictive power of the Halo EFT approach for the $3/2^+$ and
  $5/2^+$ states which are described by a neutron in a $D$-wave
  relative to the core.
\end{abstract}

\maketitle

\section{Introduction}
Halo nuclei are weakly-bound states of a few valence nucleons and a
tightly-bound core
nucleus~\cite{Zhukov:1993aw,Hansen:1995pu,Jonson:2004,Jensen:2004zz,Riisager:2012it}. They
exemplify the emergence of new degrees of freedom close to the neutron
and proton drip lines which are difficult to describe in ab initio
approaches.  Cluster models of halo nuclei are formulated directly in
the new degrees of freedom and thus take the emergence phenomenon into
account by construction, typically using a phenomenological
interaction~\cite{Descouvemont:2008zz,Schuck:2017jtw}. These models
have improved our understanding of halo nuclei significantly.
However, they cannot be improved systematically and lack a reliable
way to estimate theoretical uncertainties.

Halo effective field theory (Halo EFT) is a systematic approach to
these systems that exploits the apparent separation of scales between
the small nucleon separation energy of the halo nucleus and the large
nucleon separation energy and excitation energy of the core
nucleus~\cite{Bertulani:2002sz,Bedaque:2003wa}.  This scale separation
defines (at least) two momentum scales: a small scale $\Mlo$ and a
large scale $\Mhi$. Halo EFT provides a systematic expansion of
low-energy observables in powers of $\Mlo/\Mhi$. Predictions made in
Halo EFT can be improved systematically through the calculation of
additional orders in the low-energy expansion.
The interaction between the core and the valence nucleons is
parametrized by contact interactions tuned to reproduce a few
low-energy observables. Note that the absence of explicit pion exchange in
the interaction indicates that the approach breaks down for momenta
of the order of the pion mass.
Similar EFT approaches can be used for systems of
atoms and nucleons at low energies
\cite{Braaten:2004rn,Hammer:2010kp}.

$^{11}$Be represents the prototype of a one-nucleon halo nucleus and
thus has been considered as a test case for Halo EFT.
It has a $J^P=1/2^+$ ground state that can be described as a neutron
in an $S$-wave relative to the $^{10}$Be core. $^{11}$Be also
has a $J^P=1/2^-$ excited state which can be considered as a neutron
in a $P$-wave relative to the core.
The electric properties of the two bound states in $^{11}$Be
were studied in detail in Ref.~\cite{Hammer:2011ye} using Halo EFT.
$^{11}$Be also has a magnetic moment due to
its halo neutron \cite{Fernando:2015jyd} but there are no magnetic
transitions between the two states because of their opposite parity.
For a recent review of Halo EFT and applications to other halo nuclei
see Ref.~\cite{Hammer:2017tjm}.

Here, we will focus on the electromagnetic properties of $^{17}$C.
This nucleus is an interesting halo candidate but has not yet been
investigated using Halo EFT. Its continuum properties cannot yet be
addressed using standard ab initio methods. It is too heavy for an
approach that employs a combination of the no-core shell model (NCSM)
and the resonating group model (RGM)\cite{Navratil:2016ycn}
but it is too light to neglect center-of-mass motion
effects as is done in coupled cluster calculations. (See
Ref.~\cite{Hagen:2012rq} for a calculation of $^{40}$Ca-proton
scattering where this approximation is well justified.).
Recent calculations in the NCSM also seem
to suggest that this nucleus is too large to obtain converged results
for its spectrum \cite{Smalley:2015ngy} with the available
computational resources.  $^{17}$C has a $J^P=3/2^{+}$ ground state,
and two excited states with $J^P=1/2^{+}$ and $5/2^{+}$
\cite{Suzuki:2008zz}.  The neutron separation energy of the ground
state of about 0.7 MeV \cite{Wang:2017ame} is significantly smaller
than the excitation energy of the $J^P=0^{+}$ $^{16}$C core, which is
about 1.8 MeV \cite{Ajzenberg-Selove:1986lxy}, while the neutron
separation energies of the excited states are only of order 0.4-0.5
MeV \cite{Smalley:2015ngy} (see the level scheme in
Fig.~\ref{fig:levelschemeC17}).  This suggests that $^{17}$C may be
amenable to a description using Halo EFT with $S$- and $D$-wave
neutron-core interactions~\cite{Braun:2018hug}.

Recently, M1 transition rates from both excited states into the ground
state were measured \cite{Suzuki:2008zz,Smalley:2015ngy}. Below, we
will discuss these transition rates in the framework of Halo EFT
to leading order (LO) in the Halo EFT counting. Besides these
electromagnetic transitions, we will also consider static electric and
magnetic properties as well as neutron capture on $^{16}$C into $^{17}$C.
We will show that future experiments and/or ab initio
calculations of these quantities can provide insight in the
interaction of neutrons with $^{16}$C.

This manuscript is organized as follows: In
Sec. \ref{sec:Halo_EFT_Formalism}, we introduce the theoretical
foundations required to calculate the properties of halo nuclei with
effective field theory. After reviewing results for the charge radius
and quadrupole moment for the $S$- and $D$-wave states in
Sec. \ref{sec:static}, we calculate magnetic moments for both
states. In Sec. \ref{sec:trans_cap} we discuss E2 and M1 transitions
between the different states in $^{17}$C and calculate E1 and M1
capture reactions to the $S$- and $D$-wave states.  We end with a
summary and an outlook.

\section{Halo EFT formalism for $^{17}$C}
\label{sec:Halo_EFT_Formalism}
Our goal is to investigate the electromagnetic properties
of the halo nucleus $^{17}$C using Halo EFT.
As discussed above, $^{17}$C can be described as a weakly-bound
state of a
$^{16}$C core and a neutron. First, we need to account for the free
propagation of the core and neutron degrees of freedom. The
corresponding Lagrangian is
\begin{equation}
  \label{eq:L0}
  \mathcal{L}_0 = c^\dagger \left(i \partial_t +
    \frac{\nabla^2}{2M}\right) c
  + n^\dagger \left(i \partial_t + \frac{\nabla^2}{2m}\right) n ~,
\end{equation}
where $n$ denotes the spin-1/2 neutron field, $c$ the spin-0
core field, $m$ is the nucleon mass, and $M$ is the mass of
the $^{16}$C core. 

\begin{figure}[t]
  \centering
  \includegraphics[width=0.45\columnwidth]{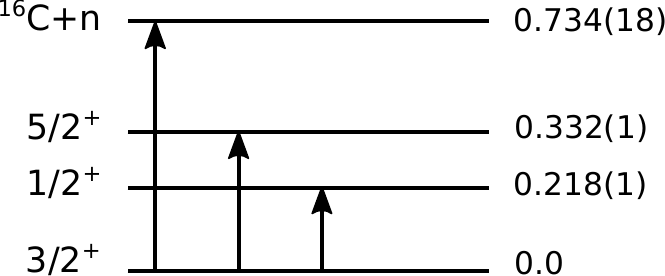}
  \caption{Level scheme of $^{17}$C showing quantum numbers $J^P$, excitation
    energies in MeV, and the $^{16}$C$~+~n$ threshold.}
  \label{fig:levelschemeC17}
\end{figure}
The first excitation of the $^{16}$C core has an energy of
$E_{{}^{16}C}^* = 1.766(10)$~MeV~\cite{Ajzenberg-Selove:1986lxy},
while the neutron separation energy of $^{16}$C is
$S_n(^{16}{\rm C})= 4.250(4)$~MeV~\cite{Wang:2017ame}.
Moreover, the neutron separation energy of $^{17}$C is
$S_n(^{17}{\rm C})= 0.734(18)$~MeV~\cite{Wang:2017ame}. This suggests
that the $J^P=3/2^{+}$ ground state of $^{17}$C can be described as
a neutron in a $D$-wave relative to the $^{16}$C core, although
the halo nature of the ground state is not commonly accepted \cite{Suzuki:2008zz,Smalley:2015ngy}.
As illustrated in Fig.~\ref{fig:levelschemeC17}, $^{17}$C also has two
excited states with
$J^P=1/2^+$ and $5/2^+$ with energies $E^*_{1/2^+} = 0.218(1)$~MeV and
$E^*_{5/2^+} = 0.332(1)$~MeV~\cite{Smalley:2015ngy}, respectively.
In Halo EFT,
these two states are described by a neutron in an $S$-wave and
$D$-wave relative to the core, respectively.
To account for these states, we define the interaction part of
the effective Lagrangian as~\cite{Braun:2018hug}
\begin{align}
\notag
  \mathcal{L} = \mathcal{L}_0 &+ d_{J,M}^\dagger \left[c_2^J \left(i \partial_t +
      \frac{\nabla^2}{2M_{nc}}\right)^2
    + \eta_2^J \left(i \partial_t + \frac{\nabla^2}{2M_{nc}}\right) + \Delta_2^J \right] d_{J,M}\\ \notag
  &- g^{J}_2 \left[d^\dagger_{J,M} \left[n \overset{\leftrightarrow}\nabla^2
      c\right]_{J,M}
    + \left[n \overset{\leftrightarrow}
      \nabla^2 c \right]^\dagger_{J,M} d_{J,M} \right]
  \\
  &+ \sigma_s^\dagger \left[\eta_0 \left(i \partial_t +
         \frac{\nabla^2}{2M_{nc}}\right) + \Delta_0 \right] \sigma_s   - g_0 \left[c^\dagger n_s^\dagger \sigma_s + \sigma_s^\dagger n_s c\right]+\ldots\ ,
\label{eq:lagrangian_dwave}
\end{align}
where $M_{nc}=M+m$ and 
$d_{J,M}$ is a $(2J+1)$-component field.
 We project on the $J=3/2$ and
 $5/2$ parts of the resonant $D$-wave interaction via
\begin{align}
\label{eq:d_tensor_L}
  \left[n \overset{\leftrightarrow}\nabla^2 c\right]_{J,M}
  &=
    \sum_{m_s
    m_l}
    \left(\left. \frac{1}{2} m_s \ 2 m_l \right\vert
    J\, M \right) \ n_{m_s} \sum_{\alpha \beta} \left(\left. 1
    \alpha \ 1 \beta \right\vert 2 m_l \right) \frac{1}{2} \left(
    \overset{\leftrightarrow}\nabla_\alpha
    \overset{\leftrightarrow}\nabla_\beta
    +  \overset{\leftrightarrow}\nabla_\beta \overset{\leftrightarrow}\nabla_\alpha \right) \ c \ ,                                                          
\end{align}
where $\alpha$ and $\beta$ denote spherical indices
and $\overset{\leftrightarrow}\nabla$ is a Galilei-invariant derivative.
The $D$-wave interaction introduces 4 low-energy constants in the
leading order (LO)
Lagrangian: $c_2^J$, $\Delta_2^J$, $g^J_2$, and $\eta_2^J = \pm
1$, but only three of them are independent at LO.
This increased number of parameters 
compared to the $S$-wave
arises from the appearance of power divergences up to 5th order in the
$D$-wave self-energy. Their renormalization requires effective range
parameters up to the shape parameter to enter at
LO~\cite{Bertulani:2002sz}. In this work, we will follow
Ref.~\cite{Braun:2018hug} and use dimensional regularization
with the power divergence subtraction scheme
(PDS) \cite{Kaplan:1998tg, Kaplan:1998we} for all practical calculations. 

The accuracy of this approach is set by the ratio of the low-momentum
scale $\Mlo$ over the high-momentum scale $\Mhi$ which for ground state
observables can be estimated as $\sqrt{S_n(^{17}{\rm C})/E_{{}^{16}C}^* }
\approx 0.64$ in our case. The expansion parameter is relatively large, and we
expect slow convergence for ground state observables. However, for the
excited states, the expansion parameter is approximately 0.5 which
leads to 50\% errors at first order and 25\% errors at second order in
the EFT expansion.

\begin{figure}[t]
  \centering
  \includegraphics[width=0.8\columnwidth]{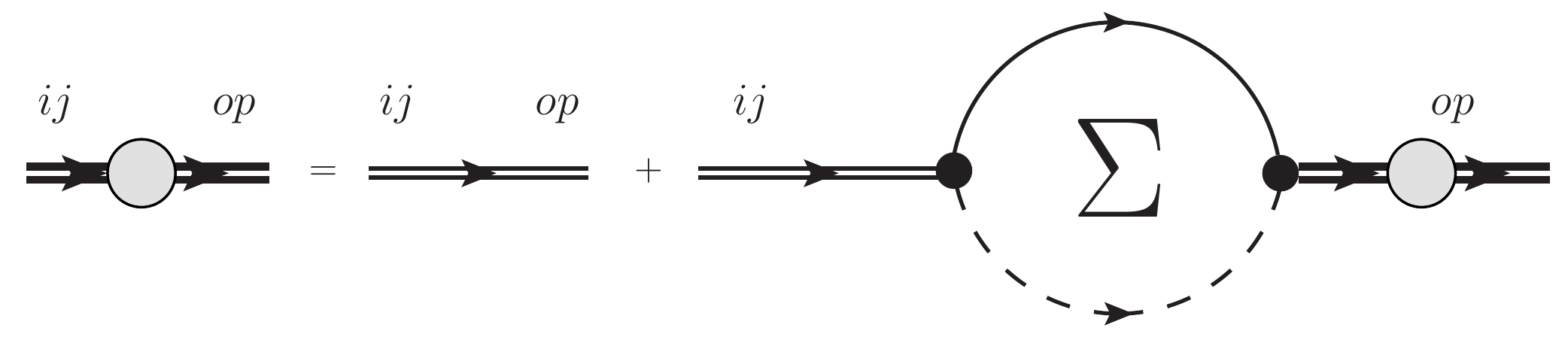}
  \caption{Diagrammatic representation of the dressed $d$-propagator.
    The dashed (solid) line denotes the core (neutron) field.
    The thin double line represents the bare $d$-propagator, while
    the thick double line with the blob is the dressed $d$-propagator.}
  \label{fig:dimer-propagator}
\end{figure}

The dressed propagators of the $\sigma$ and $d_{J,M}$ fields are obtained
by summing the bubble diagrams for the $nc$-interactions
(cf. Fig.~\ref{fig:dimer-propagator} for the $D$-wave case) to all orders.
Throughout this paper, a thick single line denotes the dressed
$\sigma$-propagator and a thick double line the dressed $d$-propagator
in all our figures.

{\em $\sigma$-propagator.}
The $\sigma$-propagator for the $S$-wave state is well known
(see, e.g., Ref.~\cite{Hammer:2011ye}) and we quote only the final result:
\begin{align}
 D_\sigma(\tilde{p}_0) &= \frac{1}{\Delta_0 + \eta_0 [\tilde{p}_0 +i\epsilon] -\Sigma_\sigma(\tilde{p}_0)}~,\\[6pt]
 \Sigma_\sigma(\tilde{p}_0) &= - \frac{g^2_0 m_R}{2 \pi} \left[i \sqrt{2 m_R \tilde{p}_0} + \mu\right]~,
\end{align}
where $\mu$ is the PDS scale~\cite{Kaplan:1998tg, Kaplan:1998we},
$m_R$ the reduced mass of the neutron-core system,
and $\tilde{p}_0=p_0 - \mathbf{p}^2/(2 M_{nc})$ is the Galilei
invariant energy.

{\em $d$-propagator.}
The dressed propagator for the $d_{J,M}$ field was computed
in Ref. \cite{Braun:2018hug}.\footnote{See also Ref.~\cite{Brown:2013zla} for
  a previous calculation using dimensional regularization with minimal
  subtraction which ignores power law divergences and
  sets $\eta_2=c_2=0$ at LO.} Since we use a
Cartesian representation of the $D$-wave, the propagator
depends on four vector indices, two in the incoming channel and two
in the outgoing channel. 
Note that Roman indices refer to Cartesian indices and Greek
ones to spherical indices.
Evaluating the Feynman diagrams
in Fig.~\ref{fig:dimer-propagator}, we obtain:
\begin{align}
\label{eq:d-wave-tensor}
& D_d(\tilde{p}_0)_{ij,op} = D_d(\tilde{p}_0) \ \frac{1}{2}\left(\delta_{io} \delta_{jp} + \delta_{ip} \delta_{jo} - \frac{2}{3} \delta_{ij} \delta_{op}\right) \ , \\
& D_d(\tilde{p}_0) = \left[\Delta_2 + \eta_2 \tilde{p}_0 + c_2 \tilde{p}_0^2 - \Sigma_d(\tilde{p}_0)\right]^{-1} \ , 
\end{align}
with the one-loop self-energy
\begin{align}
  & \Sigma_d(\tilde{p}_0) =- \frac{2}{15} \frac{m_R g_2^2}{2 \pi} \
  (2m_R\tilde{p}_0)^2
  \left[i \sqrt{2m_R\tilde{p}_0+i\epsilon} - \frac{15}{8} \mu \right] \ . 
 \label{eq:d_s-en_cutoff}
\end{align}
The term proportional to $c_2$ in \eqref{eq:lagrangian_dwave}
is required to absorb the $\mu$-dependence from the PDS scheme.
Following the arguments in Ref. \cite{Braun:2018hug}, the terms proportional to $\eta_2$, $\Delta_2$, and $g_2$
are also required to be consistent with the threshold expansion of the
scattering amplitude. In a momentum cutoff scheme, these terms absorb
the linear, cubic, and quintic power law divergences in the
cutoff~\cite{Bertulani:2002sz}.

{\em Power counting.} The canonical power counting for the
$\sigma$-propagator representing a shallow $S$-wave state was given in
Refs.~\cite{vanKolck:1997ut,vanKolck:1998bw,Kaplan:1998tg,Kaplan:1998we}.
It implies $\gamma_0 \sim  1/a_0 \sim \Mlo$ and $r_0 \sim 1/\Mhi$,
where $\gamma_0= \sqrt{2 m_R (S_n(^{17}\mathrm{C})-E^*_{{1/2}^+})}$ is the binding momentum of the
$S$-wave state and $r_0$ the effective range.
As a result, $r_0$ enters at NLO in the expansion in $\Mlo/\Mhi$.

The power counting for partial waves beyond the $S$-wave is more
complicated and different scenarios
have been proposed~\cite{Bertulani:2002sz,Bedaque:2003wa,Braun:2018hug}.
We look for a scenario that exhibits the minimal number of fine tunings
consistent with the scales of the system.
Bedaque et al.~\cite{Bedaque:2003wa}  
suggested for the $P$-wave case that $a_1 \sim 1/(\Mlo^2 \Mhi)$ and
$r_1 \sim \Mhi$, where higher ERE parameters scale with the appropriate
power of $\Mhi$ given by dimensional analysis.
This power counting is adequate for the excited
state of $^{11}$Be~\cite{Hammer:2011ye}.
It requires only one fine-tuned constant in $\mathcal{L}$ instead of
two as proposed in Ref.~\cite{Bertulani:2002sz} where both $a_1$ and $r_1$
scale with appropriate powers of $\Mlo$.
In Ref.~\cite{Bedaque:2003wa}, the power counting was also generalized to $l>1$. 
However, we employ a different power counting with a minimal number of
fine tunings for $l=2$ as proposed in Ref.~\cite{Braun:2018hug}.
In the case of the
$d$-propagator, \eqref{eq:d-wave-tensor}, two out of three ERE
parameters need to be fine-tuned because $a_2 \sim 1/(\Mlo^4
\Mhi)$ and $r_2 \sim \Mlo^2
\Mhi$ are both unnaturally large, while $\mathcal{P}_2 \sim
\Mhi$.  Higher ERE terms are suppressed by powers of $\Mlo/\Mhi$.
Thus, the relevant fit-parameters in our EFT at LO
are $\gamma_0$, $\gamma_2$, $r_2$, and $\mathcal{P}_2$,
where $\gamma_2 = \sqrt{2m_R S_n(^{17}\mathrm{C})}$
is the binding momentum of the $^{17}$C ground state, while $r_2$ and
$\mathcal{P}_2$ denote the $D$-wave effective range and shape parameter,
respectively. For the $5/2^+$ excited state, the binding
momentum is $\gamma_{2'} =\sqrt{2m_R (S_n(^{17}\mathrm{C})-E^*_{5/2^+})}$,
while $r_{2'}$, $\mathcal{P}_{2'}$ are the corresponding effective range parameters.

The corresponding wave function renormalization constants
for the $1/2^+$, $3/2^+$, and $5/2^+$ states
at LO are:
\begin{align}
\label{eq:normalizationSD}
& Z_\sigma = \frac{2\pi}{m_R^2 g_0^2} \ \gamma_0 \ , \qquad
 Z_d^{3/2} = -\frac{15\pi}{m_R^2 g^2_2} \ \frac{1}{r_2 + \mathcal{P}_2 \gamma_2^2} \ , \qquad
 Z_{d'}^{5/2} = -\frac{15\pi}{m_R^2 g^2_{2'}} \ \frac{1}{r_{2'} + \mathcal{P}_{2'} \gamma_{2'}^2} \ ,
\end{align}
respectively.
At NLO, $Z_\sigma$ is modified by a factor $(1+\gamma_0 r_0)$.
The constants $Z_d^{3/2}$ and $Z_{d'}^{5/2}$ are only required at LO for our calculations.

\section{Static electromagnetic properties of $^{17}$C}
\label{sec:static}
We first consider the static electromagnetic properties of
$^{17}$C. These are usually easier to measure experimentally
than dynamical properties. They can
also be calculated in ab initio approaches that provide the wave
functions of the involved states. In particular, we will consider the
charge radii and magnetic moments of the $^ {17}$C states.
It is convenient to calculate all form factors in the Breit frame where the
photon transfers no energy,
$q=(0,\boldsymbol{q})$, and to choose the photon to be moving in the
$\hat{z}$ direction $\boldsymbol{q} = |\boldsymbol{q}| \hat{z}$.

\subsection{Charge radii}
\label{sec:charge-radius}
The form factor of a general $S$-wave one-neutron halo nucleus was
calculated in Ref.~\cite{Hammer:2011ye}.
The electric charge radius of the $S$-wave state at NLO is given by:
\begin{align}
  \braket{r_E^2}^{(\sigma)} = \frac{f^2}{2 \gamma_0^2} (1+r_0\gamma_0)\ .
   \label{eq:rE-swave}
\end{align}
where $f=m_R/M$ is a mass factor.  The LO result can be obtained by
setting $r_0=0$ in Eq.~(\ref{eq:rE-swave}).
At next-to-next-to-leading order (NNLO) a counterterm
related to the radius of the core contributes.
In the standard power counting, the factors
of $f$ are counted as ${\cal O}(1)$, although they can become rather small
for large core masses. As a consequence, the counterterm contribution is
enhanced numerically. Up to NLO, one can interpret the Halo EFT result as
a prediction for the radius relative to the core~\cite{Hammer:2011ye}.

Using the measured one-neutron separation energy of the $1/2^{+}$ state, we
obtain for the charge radius of the excited $S$-wave
state of $^{17}$C relative to the charge radius of  $^{16}$C at LO:
\begin{equation}
\label{eq:charge_radius}
  \braket{r_E^2}^{1/2^{+}}_{^{17}C} - \braket{r_E^2}_{^{16}C} = 0.074 \ \text{fm}^2 \ , 
\end{equation}
where the error from NLO corrections is about 50\%. To make a
numerical prediction for the full charge radius of $^{17}$C, we have
to add the charge radius of $^{16}$C, $\braket{r_E^2}_{^{16}C}$, to
our result. For this purpose, we use the point-proton radius $R_p$ from
Ref.~\cite{Kanungo:2016tmz} and the formula for the charge radius from
Ref.~\cite{Mueller:2008bj},
including the Darwin-Foldy term and the neutron charge radius
as corrections, to obtain
$\sqrt{\braket{r_E^2}^{1/2^{+}}_{^{17}C}} = \sqrt{\left(R_p^{^{16}C}\right)^2
  +r_p^2 + \frac{3}{4 m^2} + \frac{N}{Z} r_n^2 + 0.074 \text{ fm}^2} =
2.53(5)$ fm. Here we have used the proton, $r_p = 0.875$ fm, and neutron
charge radii, $r_n^2 = -0.116$ fm$^2$ \cite{Yao:2006px}, and $N = 11$
($Z = 6$) denotes the number of neutrons (protons) of $^{17}$C. The error
bar includes both the experimental and the Halo EFT uncertainties.

To date, there is no experimental data for the charge radius of the
$1/2^{+}$ excited state to compare with.  As a consistency check, we
compare with the experimental value for the $3/2^+$ ground state of
$^{17}$C extracted in Ref.~\cite{Kanungo:2016tmz},
$\sqrt{\braket{r_E^2}^{3/2^+}_{^{17}C}} = 2.54(4)$ fm, which is very
close to our result for the $1/2^{+}$ excited state. Note that the difference
between the charge radius of $^{17}$C and $^{16}$C is smaller than the
experimental error from Ref. \cite{Kanungo:2016tmz} for this quantity.

The charge radius of a $D$-wave state has recently been calculated in
Ref.~\cite{Braun:2018hug} at LO and yields:
\begin{align}
\label{eq:r_E_LO}
 \braket{r_E^2}^{(d)} \ =  -\frac{6 \tilde{L}_{C0E}^{(d) \text{ LO}}}{r_2 + \mathcal{P}_2 \gamma_2^2} \ .
\end{align}
Here, the counterterm $\tilde{L}_{C0E}^{(d) \text{ LO}}$ already
contributes at LO while the loop contribution is suppressed.

For the $D$-wave state, we also find a quadrupole moment which yields
at LO:
\begin{align}
\label{eq:Q_LO}
  \mu_Q^{(d)} &=
                \frac{40 \tilde{L}_{C02}^{(d)\text{ LO}}}
                {3 \left(r_2 + \mathcal{P}_2 \gamma_2^2 \right)} \ ,
\end{align}
where another counterterm enters at LO.
Both $D$-wave observables have the same denominator of effective range
parameters $(r_2 +\mathcal{P}_2 \gamma_2^2)$ which is related to the Asymptotic Normalization Coefficient (ANC) of the $D$-wave state,
$A_2 = \sqrt{2 \gamma_2^4/(-r_2 -\mathcal{P}_2 \gamma_2^2)}$. Similar to the correlation between
$\mu_Q^{(d)}$ and B(E2) in Ref.~\cite{Braun:2018hug}, we find a
smooth correlation between $\braket{r_E^2}^{(d)}$ and $\mu_Q^{(d)}$:
\begin{align}
  \mu_Q^{(d)} = -\frac{20}{9}
  \frac{\tilde{L}_{C02}^{(d)\text{ LO}}}{\tilde{L}_{C0E}^{(d)\text{ LO}}}
  \braket{r_E^2}^{(d)} \ ,
\end{align}
which implies that ab initio calculations with different phaseshift-equivalent
interactions should show a linear correlation between
the quadrupole moment and the charge radius.
%
%
\subsection{Magnetic moments}
\label{sec:magnetic-moments}
\begin{figure}
  \centering
  \includegraphics[width=0.89\textwidth]{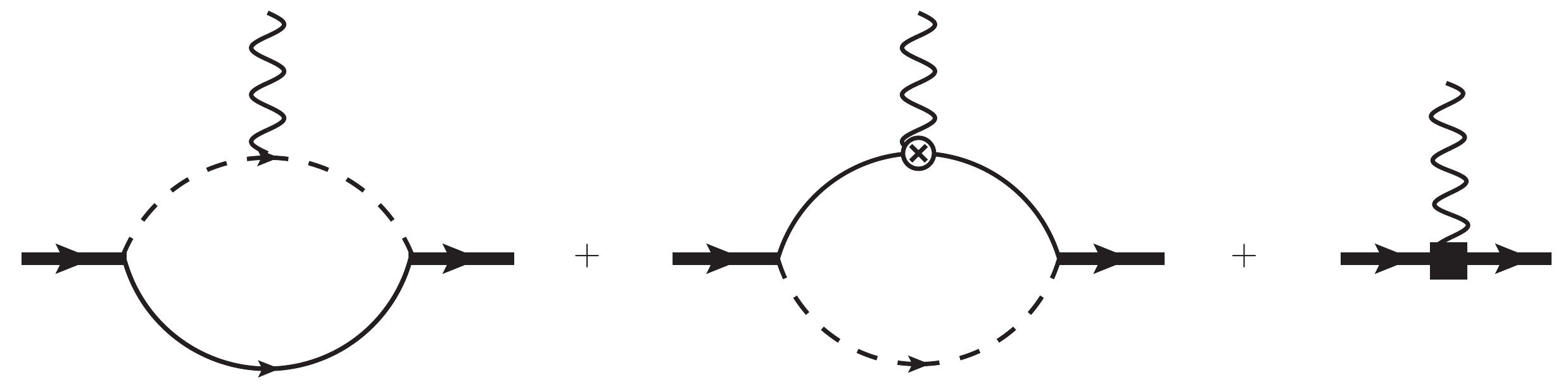}
  \caption{Diagrams contributing to the magnetic moment. The first
    diagram is the coupling of a vector photon to the charge of the
    core arising from minimal substitution in the Lagrangian.  The
    second diagram displays a vector photon coupling to the magnetic
    moment of the neutron. The last diagrams shows a two-body current. 
    The thick solid line denotes the dressed $\sigma$-propagator.}
  \label{fig:diagramsM1momentGammaV}
\end{figure}
The magnetic properties of shallow bound states are predominantly
determined by the magnetic moments of its degrees of freedom. The
magnetic moment of a single particle is introduced into the Lagrangian
through an additional {\it magnetic} one-body operator
\cite{Chen:1999tn,Fernando:2015jyd}. An additional
counterterm enters via a two-body current. Assuming a spin-0
core, the effective Lagrangian is
\begin{align}
\label{eq:L_magnetic}
  \mathcal{L}_{M} = \kappa_n \mu_N n^\dagger \boldsymbol{\sigma \cdot B} n
  + 2 \mu_N L_M^J \Phi^\dagger \boldsymbol{S_J \cdot B} \Phi ~,
\end{align}
where $\Phi$ is a place holder for the relevant auxiliary field
($\sigma_s$, $\pi_s$, $d_{J,M}$, ...), $\boldsymbol{S_J}$ is the
corresponding spin matrix for spin $J$, $\mu_N$ denotes the nuclear
magneton, and $L^J_M$ the coupling constant for the magnetic two-body
current. For the neutron anomalous
magnetic moment we use $\kappa_n = -1.91304$.

\subsubsection{Magnetic moment of the $1/2^{+}$ state}
\label{sec:magetic-moment-swave}

We reproduce the results obtained by Fernando {\it et al.}
\cite{Fernando:2015jyd}, who calculated electromagnetic form factors
for $S$-wave states of one-neutron halo nuclei. Up to NLO, only the two last
diagrams in Fig.~\ref{fig:diagramsM1momentGammaV} contribute to the
magnetic form factor in the Breit frame:
\begin{align}
  \frac{e Q_c}{2 M_{nc}} G_M(q^2) &= Z_\sigma \mu_N \left(g_0^2 \kappa_n \frac{m m_R}{\pi q} \arctan\left[\frac{q m_R}{2 m \gamma_0}\right] + L_M^\sigma \right)~,
\end{align}
with
\begin{align}
  Z_\sigma = \frac{2\pi \gamma_0}{m_R^2 g_0^2} (1+r_0\gamma_0)~, \qquad \text{and we define} \qquad
  \tilde{L}_M^\sigma = \frac{2 \pi L_M^\sigma}{m_R^2 g_0^2} \ .
\end{align}
The magnetic moment $\kappa_\sigma$ is obtained by
evaluating the form factor at $q^2=0$:
\begin{align}
\kappa_\sigma &= \frac{e Q_c}{2 M_{nc}} G_M(0) =\left( \kappa_n + \tilde{L}_M^\sigma \gamma_0 \right) (1+r_0\gamma_0)~,
\end{align}
where $\kappa_\sigma$ is given in units of $\mu_N$.
Naive dimensional analysis with rescaled fields
$[\tilde{\sigma}] = 2$ \cite{Hammer:2011ye} determines the scaling
of the counterterm
$\tilde{L}_M^\sigma \sim \Mhi^{-1}$. As a consequence,  $\tilde{L}_M^\sigma$
contributes at NLO.
At LO, the magnetic moment of the
$1/2^{+}$ state is thus given by the magnetic moment
of the neutron, $\kappa_n$.

\subsubsection{Magnetic moments of the $3/2^{+}$ and $5/2^{+}$ states}
\label{sec:magetic-moment-dwave}
In the case of the $D$-wave, the only contribution to the magnetic
moment at LO is the two-body current in Eq. \eqref{eq:L_magnetic},
which corresponds to the last diagram in
Fig. \ref{fig:diagramsM1momentGammaV}, and we obtain:
\begin{align}
  \frac{e Q_c}{2 M_{nc}} G_M(q^2) &= Z_d \mu_N L_M^d = -\frac{\mu_N \tilde{L}_M^d}{r_2 + \mathcal{P}_2 \gamma_2^2} \ ,
\end{align}
with
\begin{align}
  Z_d = -\frac{15\pi}{m_R^2 g_2^2} \frac{1}{r_2 + \mathcal{P}_2 \gamma_2^2}~, \qquad \text{and} \qquad
  \tilde{L}_M^d = \frac{15 \pi L_M^d}{m_R^2 g_2^2} \ .
\end{align}
This yields for the magnetic form factor at LO:
\begin{align}
  \kappa_d = - \frac{ \tilde{L}_M^d}{{r_2 + \mathcal{P}_2 \gamma_2^2}} ~,
\end{align}
where $\kappa_d$ is again given in units of $\mu_N$.
Beyond LO we also need to consider the two loop diagrams in
Fig. \ref{fig:diagramsM1momentGammaV}. Therefore, we require
additional counterterms to renormalize the corresponding
divergences. This makes predictions even harder, and for that reason,
we do not calculate the NLO contribution to the magnetic form factors
for the $D$-wave state explicitly.

In general, the magnetic moment of the
$D$-wave states will thus differ significantly from the magnetic moment of
the neutron since $\kappa_n$ is a NLO contribution.

\section{Electromagnetic transitions and capture reactions of $^{17}$C}
\label{sec:trans_cap}

\subsection{E2 transitions}
The ground state and the two excited states of $^{17}$C have positive
parity and differ at most by 2 units in total angular momentum. All
states can therefore be connected by E2 transitions.

The transition strength for $S \rightarrow D'$ has been calculated at LO in
Ref.~\cite{Braun:2018hug} for the transition:
\begin{align}
\label{eq:E2_LO}
 \text{B(E2: $1/2^+ \to 5/2^+$)} &= -\frac{4}{5 \pi} \frac{Z_{eff}^2 e^2}{r_{2'} +\mathcal{P}_{2'} \gamma_{2'}^2} \ \gamma_0 \ \left[ \frac{3\gamma_0^2 + 9\gamma_0\gamma_{2'} + 8\gamma_{2'}^2}{(\gamma_0 + \gamma_{2'})^3} \right]^2 ,
\end{align}
where the effective charge for $^{17}$C, $Z_{eff} = (m/M_{nc})^2 Q_c \approx
0.021$~\cite{Typel:2004us}, comes out of the calculation
automatically. At NLO, there is an
unknown short-range contribution that enters via a counterterm.

For the transition strength B(E2: $1/2^+ \to 3/2^+$), we get the same
result for the amplitude but with different Clebsch Gordan coefficients
(leading to a relative factor of $3/2$) and the appropriate binding momentum
and renormalization constant for the $3/2^+$ ground state:
\begin{align}
\label{eq:E2_LOx}
 \text{B(E2: $1/2^+ \to 3/2^+$)} &=- \frac{8}{15 \pi} \frac{Z_{eff}^2 e^2}{r_2 +\mathcal{P}_2 \gamma_2^2} \ \gamma_0 \ \left[ \frac{3\gamma_0^2 + 9\gamma_0\gamma_2 + 8\gamma_2^2}{(\gamma_0 + \gamma_2)^3} \right]^2 .
\end{align}

Following the approach in Ref.~\cite{Braun:2018hug}, we can also
calculate the E2 transition for $D \to D'$. However, we do not display
the result here since the relevant diagram diverges cubically and,
therefore, additional counterterms are required for this observable
already at LO.

\subsection{M1 transitions}
\subsubsection{S $\rightarrow$ D}
We will first consider the M1 transition strength from the $3/2^+$
ground state ($D$-wave) to the first excited $1/2^+$ state ($S$-wave) in
$^{17}$C since it was measured in Refs.~\cite{Suzuki:2008zz,Smalley:2015ngy}.
The experimental result is small compared
with typical M1 transition strengths in nuclei, {\it i.e.}
$\text{B(M1: $1/2^+ \to 3/2^+$)} = 1.04^{+0.03}_{-0.12} \times 10^{-2}
\mu_N^2$ \cite{Smalley:2015ngy} or $0.58 \times 10^{-2}$
W.U. expressed in Weisskopf units.

In the neutron-core picture of Halo EFT, the M1 transition from a $D$-wave
to an $S$-wave state is forbidden for one-body currents which is in agreement
with the experimental suppression of the transition.
The non-zero transition strength can only be accounted for by
a two-body current which takes  short-ranged (core) physics into account.
We therefore add the gauge-invariant counterterm
\begin{align}
 \mathcal{L}_{M} = -\mu_N L_{M1}^{\sigma d} \sigma_m^\dagger d_{m'} \left(\frac{1}{2} m 1 i \bigg| \frac{3}{2} m' \right) B_i \ .
\end{align}
By rescaling the fields to absorb unnaturally large coupling
constants, leading to $[\tilde{\sigma}] = 2$, $[\tilde{d}] = 0$,
and using naive dimensional analysis for the rescaled fields
\cite{Beane:2000fi}, we find
$L_{M1}^{\sigma d} \sim \Mhi l_{M1}^{\sigma d} g_0 g_2 m_R^2$
with $l_{M1}^{\sigma d}$ of order one. To obtain the magnetic
transition amplitude we calculate the vertex function
\begin{align}
 \Gamma_{m m' i} = \left(\frac{1}{2} m 1 i \bigg| \frac{3}{2} m' \right) \mu_N \tilde{L}_{M1}^{\sigma d} \epsilon_{ijk} k_j ~,
\end{align}
with $\tilde{L}_{M1}^{\sigma d} = \frac{\sqrt{30}\pi}{m_R^2 g_0 g_2} L_{M1}^{\sigma d}$.
If we consider the case $m = -m' = \pm 1/2$ and choose the photon to be traveling in $\hat{z}$ direction, we find
\begin{align}
 \bar{\Gamma}_{\pm \mp, \mp1} = \mp \frac{\mu_N}{\sqrt{3}} \tilde{L}_{M1}^{\sigma d} \omega ~.
\end{align}
This yields for the M1 transition strength:
\begin{align}
  \text{B(M1: $1/2^+ \to 3/2^+$)} =
  \frac{3}{4\pi} \left( \frac{\bar{\Gamma}_{\pm \mp, \mp1}}{\omega}\right)^2 =
  -\frac{1}{4\pi} \frac{\gamma_0}{r_2 +\mathcal{P}_2 \gamma_2^2}
  \left( \tilde{L}_{M1}^{\sigma d} \right)^2 \mu_N^2 ~.
  \label{eq:M1x}
\end{align}
Moreover, combining Eqs.~(\ref{eq:M1x}) and (\ref{eq:E2_LOx}),
we find a correlation between B(E2) and B(M1):
\begin{align}
  \text{B(E2: $1/2^+ \to 3/2^+$)} = \frac{32}{15} \frac{Z_{eff}^2 e^2}{\left(\tilde{L}_{M1}^{\sigma d} \right)^2 \mu_N^2} \ \left[ \frac{3\gamma_0^2 + 9\gamma_0\gamma_2 + 8\gamma_2^2}{(\gamma_0 + \gamma_2)^3} \right]^2 \text{B(M1: $1/2^+ \to 3/2^+$)} ~.
\end{align}
If we use the experimental result for B(M1: $1/2^+ \to 3/2^+$)
$= 1.04^{+0.03}_{-0.12} \times 10^{-2} \mu_N^2$ and employ naive
dimensional analysis for the counterterm
$\tilde{L}_{M1}^{\sigma d} \sim \Mhi \approx 0.28 \text{ fm}^{-1}$, we
can make a rough prediction for B(E2),
\begin{align}
  \text{B(E2: $1/2^+ \to 3/2^+$)} \approx 3 \times 10^{-2} \ e^2 \text{fm}^4\,
  .
\end{align}

Moreover, we can compare the M1 and E2 transition strengths for
$^{17}$C if we look at the transition rates \cite{Greiner:1996nuc},
\begin{align}
  T(R \lambda) = \frac{8\pi (\lambda +1)}{\lambda [(2\lambda+1)!!]^2} \omega^{2\lambda+1} B(R \lambda)~,
\end{align}
that have, in contrast to B(M1) and B(E2), the same units.  Here $R$
stands for E or M, $\lambda$ denotes the order of the
transition and $\omega$ defines the photon energy which, in this case,
is $0.218$ MeV (cf.~Fig.~\ref{fig:levelschemeC17}). Using the naive
dimensional analysis result for
$\tilde{L}_{M1}^{\sigma d}$ from above we find:
\begin{align}
  \frac{T(E2)}{T(M1)} = \frac{32 \omega^2}{125} \frac{Z_{eff}^2 e^2}{\left(\tilde{L}_{M1}^{\sigma d} \right)^2 \mu_N^2} \ \left[ \frac{3\gamma_0^2 + 9\gamma_0\gamma_2 + 8\gamma_2^2}{(\gamma_0 + \gamma_2)^3} \right]^2 \approx 1\times 10^{-5}~,
\end{align}
which implies that the M1 transition strongly dominates over E2 for $^{17}$C.

\subsubsection{D' $\rightarrow$ D}
\begin{figure}[t]
	\centering
		\includegraphics[width=0.99\textwidth]{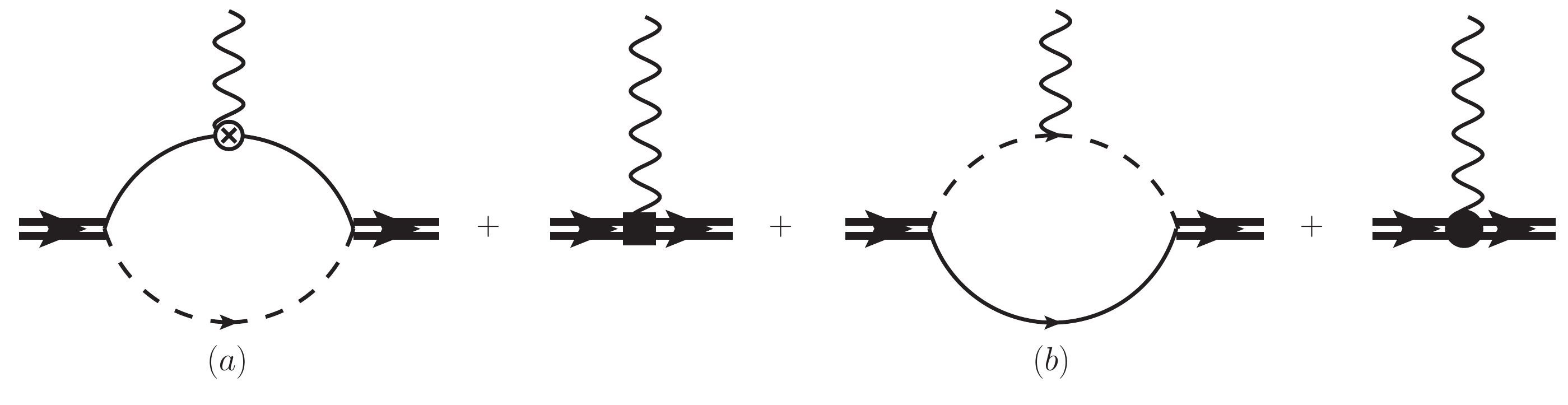}
                \caption{Relevant diagrams for the M1 transition. In the
                  diagram (a) a vector photon couples to the magnetic
                  moment of the neutron and in (b) to the
                  electric charge of the core. In the two remaining diagrams
                  the photon couples directly to the $D$-wave dimers. 
                  For a more detailed description of the lines, see Fig. \ref{fig:dimer-propagator}.}
	\label{fig:diagramsM1trans}
\end{figure}
The M1 transition strength from the $3/2^+$ ground state ($D$-wave) to
the second excited $5/2^+$ state ($D'$-wave) in $^{17}$C was also
measured in Ref.~\cite{Suzuki:2008zz}: $\text{B(M1:
  $5/2^+ \to 3/2^+$)} = 7.12_{-0.96}^{+1.27} \times 10^{-2} \mu^2_N$.
Compared to the $D \to S$-state M1 transition strength, it is around
one order of magnitude larger.
This is in agreement with the fact that M1 transitions are allowed for
neutron-core systems with one-body currents by the usual selection
rules. We calculate both loop diagrams in
Fig. \ref{fig:diagramsM1trans} and find that we need additional
counterterms to absorb all divergences. Moreover, we obtain results
for the M3 and M5 transition.
We find that two different counterterms are needed for the M1
transition and also two for the M3 transition.

In the following, we
concentrate the discussion on the M1 transition. In this case, the two
counterterms are given by:
\begin{align}
  \mathcal{L}_{M} = - L^{dd'}_{M1a} \mu_N d^\dagger_{ij} d'_{ij} \sigma_k^{m_s m_{s'}} B_k \
  - L^{dd'}_{M1b} \mu_N d^\dagger_{ij} \boldsymbol{\nabla}\cdot{\bf A} d'_{ij} \ .
\end{align}
The first counterterm is needed to renormalize the scale dependence
from diagram (a) with the magnetic photon coupling to the neutron
and the second one renormalizes the scale for the vector photon
coupling in diagram (b), respectively.  For the calculation it is
convenient to define:
\begin{align}
  \tilde{L}_{M1a}^{dd'} &= \frac{15 \pi}{m_R^2 g_{2'} g_2} L^{dd'}_{M1a} + \frac{15}{4} \left( \gamma_2^2+\gamma_{2'}^2 \right) \kappa_n \mu \ , \\
 \tilde{L}^{dd'}_{M1b} &= \frac{15 \pi}{m_R^2 g_{2'} g_2} L^{dd'}_{M1b} + \frac{15}{4} \left( \gamma_2^2+\gamma_{2'}^2 \right) \frac{m_R Q_c}{M} \mu \ ,
\end{align}
where $\mu$ is the PDS scale. 

Again, the photon has four-momentum $k = (\omega, \boldsymbol{k})$, and its
polarization index is denoted by $\nu$. The computation of both diagrams
yields a vertex function $\Gamma_{m m'\nu}$ , where $m$ is the total
angular momentum projection of the $3/2^+$ state and $m'$ denotes the
spin projection of the $5/2^+$ state. We compute the vertex function
with respect to the specific components of the $D$-wave interaction:
\begin{align}
 \Gamma_{m m' \nu} = \sum_{\alpha\beta\delta\eta m_l m_l' m_s m_s'} {\left( \frac{1}{2} m_s 2 m_l \left| \frac{3}{2} m \right. \right) \left( 1 \alpha 1 \beta \left| 2 m_l \right. \right) \left( \frac{1}{2} m_s' 2 m_l' \left| \frac{5}{2} m' \right. \right)\left( 1 \delta 1 \eta  \left| 2 m_l' \right. \right) \tilde{\Gamma}_{\alpha\beta\delta\eta \nu}} \ .
\end{align}
We calculate the irreducible vertex in Coulomb gauge so that we
have $\boldsymbol{k} \cdot \boldsymbol{\epsilon} = 0$ for real
photons. Additionally, we choose
$\boldsymbol{k} \cdot \boldsymbol{p} = 0$, where $\boldsymbol{p}$
denotes the incoming momentum of the $D$-wave state.  As a result, the
space-space components of the vertex function in Cartesian coordinates
for the left diagram can be written as:
\begin{align}
  \tilde{\Gamma}_{ijopk} =
  \Gamma_M^{(a)} \epsilon_{abk} \sigma_{a}^{m_s m_{s'}} k_b \left(\frac{\delta_{io} \delta_{jp} + \delta_{ip} \delta_{jo}}{2} - \frac{1}{3} \delta_{ij} \delta_{op}\right) \ ,
\end{align}
and for the right one:
\begin{align}
  \tilde{\Gamma}_{ijopk} =
  \Gamma_M^{(b)} p_k  \left(\frac{\delta_{io} \delta_{jp} + \delta_{ip} \delta_{jo}}{2} - \frac{1}{3} \delta_{ij} \delta_{op}\right) + \Gamma_{E2} \left[k_i \left(\frac{\delta_{j p} \delta_{k o}+\delta_{j o} \delta_{k p}}{2} - \frac{1}{3} \delta_{j k} \delta_{o p}\right) + \cdots \right] \ .
\end{align}
In the left diagram, the photon couples to the spin of the neutron and
we get a spin flip $m_s \neq m_s'$. In the case of the right diagram
there is no spin flip so that $m_s=m_{s'}$. By choosing the photon to
be traveling in $\hat{z}$ direction it follows from the tensor
structure of $\tilde{\Gamma}_{ijop\nu}$ that $m_l = m_l'$ and
$\nu \neq 0$. For the case that $m=\pm1/2=-m'$ we get:
\begin{align}
  -\Gamma_{-+,1} = \Gamma_{+-,-1} = \frac{\sqrt{6}}{5} \Gamma_M^{(a)} \sqrt{2} \omega \ ,
\end{align}
and for $m=m'$ we get $0$ for all possible values.
This yields for the B(M1: $3/2^+ \to 5/2^+$) transition:
\begin{align}
\notag
  \text{B(M1: $3/2^+ \to 5/2^+$)} &= \frac{3}{4 \pi} \left(\frac{\Gamma_{+-,-1}}{\omega}\right)^2 = \frac{9}{25\pi} \left(\frac{\bar{\Gamma}_{M}^{(a)} \omega}{\omega}\right)^2 \\
  &= \frac{9 \mu_N^2}{25\pi} \frac{1}{r_2 + \mathcal{P}_2\gamma_2^2} \frac{1}{r_{2'} + \mathcal{P}_{2'}\gamma_{2'}^2} \left[ \tilde{L}^{dd'}_{M1a} + \frac{2 \gamma _{2'}^4 \kappa_n}{ \left(\gamma _{2'}+\gamma _2\right)} + 2\kappa_n\left( \gamma _2
   \gamma _{2'}^2 + \gamma _2^3 \right)  \right]^2 \ ,
\end{align}
with the renormalized, irreducible vertex $\bar{\Gamma}_M = \sqrt{Z_d Z_{d'}} \Gamma_M$.
By rescaling the fields, $[\tilde{d}] = [\tilde{d}'] = 0$, and using
dimensional analysis we find that the counterterm scales as
$L^{dd'}_{M1a} \sim \Mhi^3 l^{dd'}_{M1a} g_2 g_{2'} m_R^2$ with
$l^{dd'}_{M1a}$ of order one. In contrast, the contribution from the
loop scales as $\Mlo^3$ which means that in LO only the counterterm
contributes to the M1 transition and the loop diagram is suppressed by
$(\Mlo/\Mhi)^3$.  Thus the M1 transition is strongly dominated by
short-range physics.

\subsection{E1 neutron capture on $^{16}$C}

\subsubsection{E1 capture into the $1/2^{+}$ state}
\begin{figure}[t]
  \centering
  \includegraphics[width=0.49\textwidth]{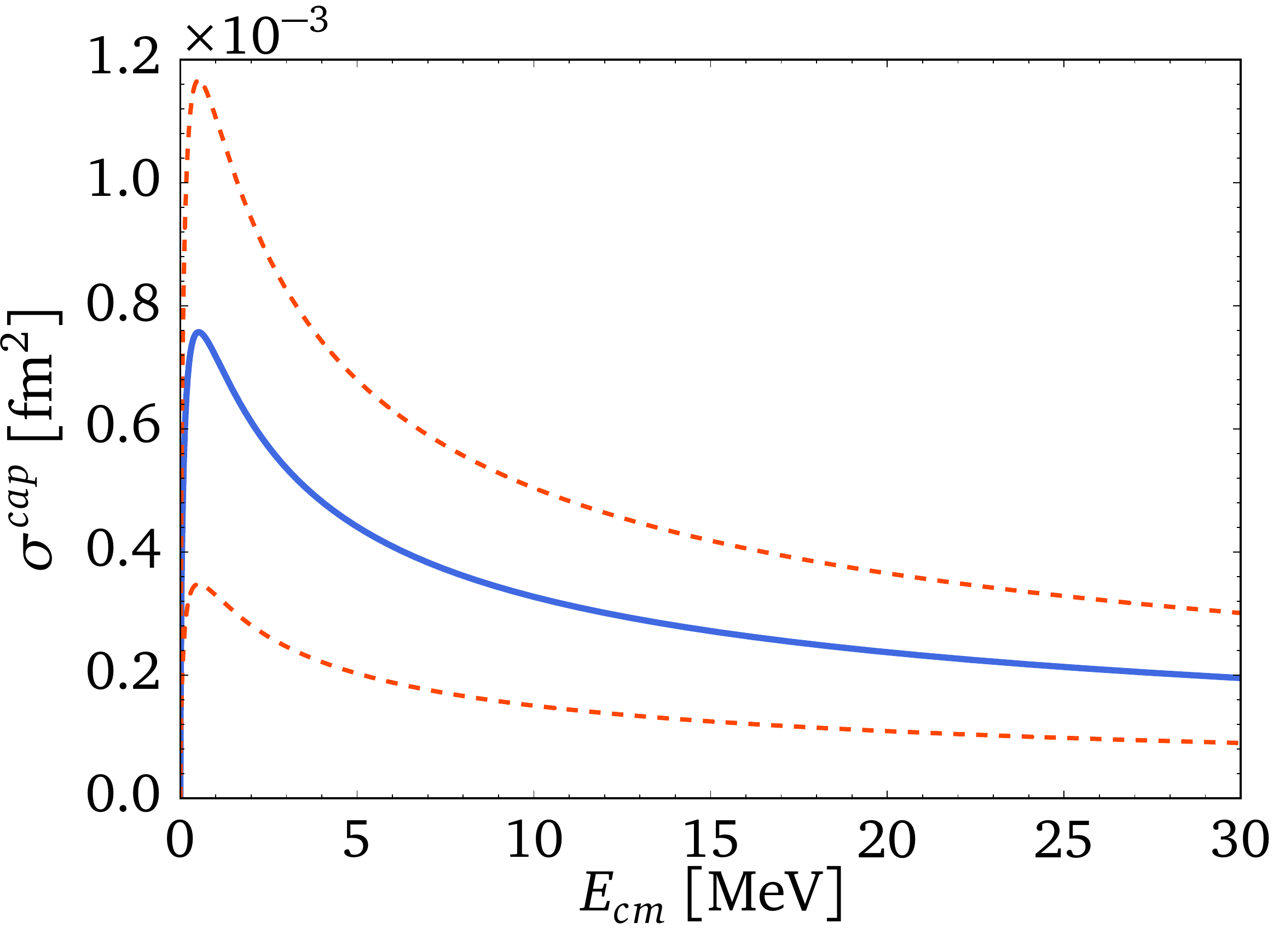}
  \includegraphics[width=0.49\textwidth]{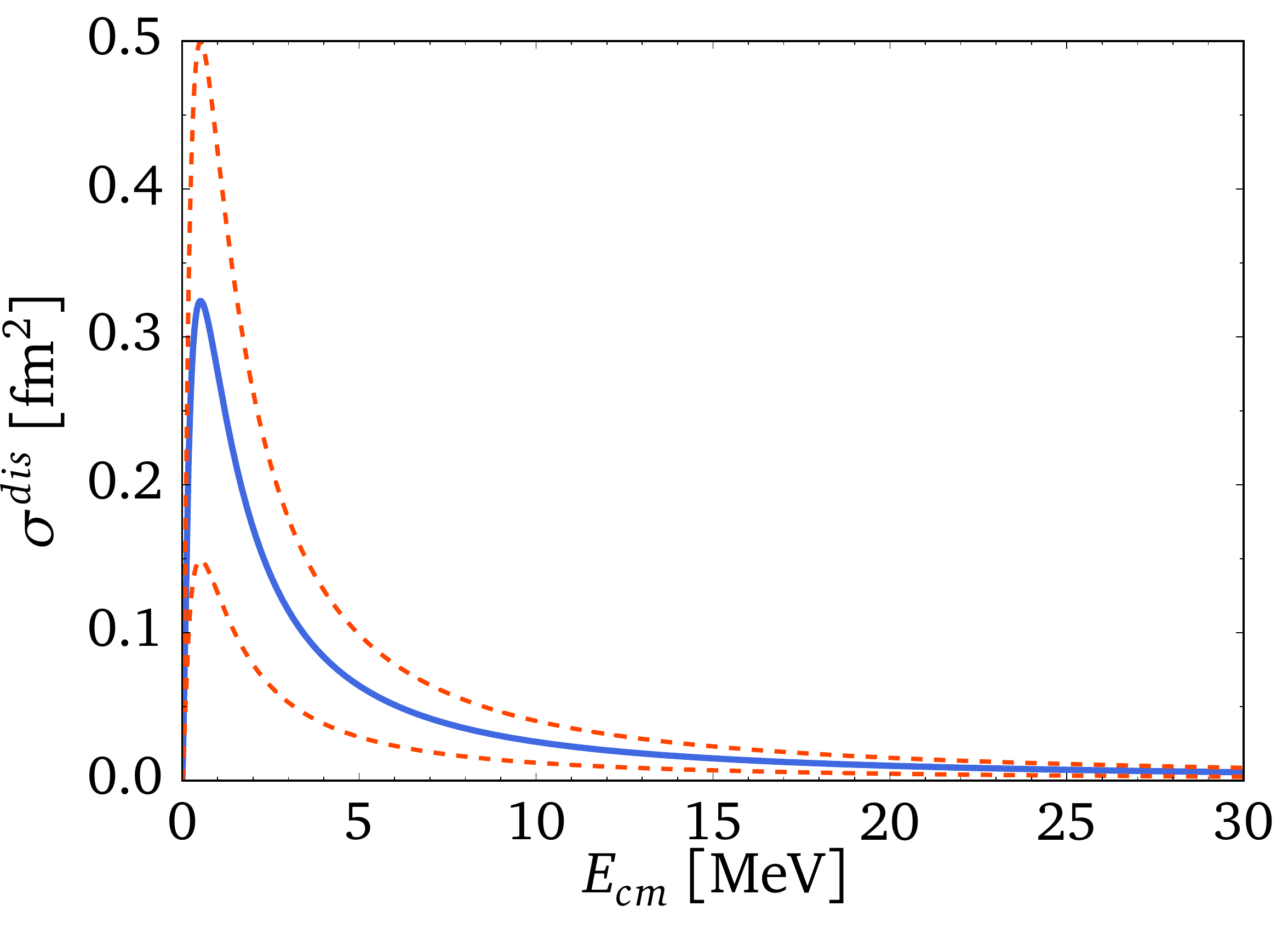}
  \caption{Left panel: E1 capture cross section into $^{17}$C as a
    function of the center-of-mass energy $E_{cm}$. Right panel: E1
    photodissociation cross section as a function of $E_{cm}$. The solid (blue) line denotes the LO result and the dashed (red) lines show an estimate of the NLO corrections.} 
  \label{fig:E1captureCrossSectionC17swave}
\end{figure}

E1 capture proceeds dominantly through the vector coupling of the
photon to the halo core. The corresponding leading order operator is
generated through minimal substitution in Eq.~\eqref{eq:lagrangian_dwave}.
\begin{figure}[t]
  \centering
  \includegraphics[width=0.3\textwidth]{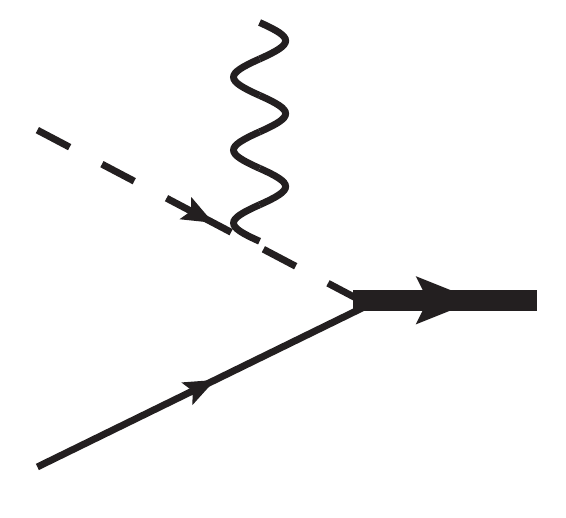}
  \caption{Relevant diagram contributing to the E1 capture amplitude to $S$-wave
    states at LO. For a more detailed description of the lines, see Fig. \ref{fig:dimer-propagator} and \ref{fig:diagramsM1momentGammaV}. }
  \label{fig:diagramsE1captureLO}
\end{figure}
The diagram that contributes at LO to this process is shown in
Fig.~\ref{fig:diagramsE1captureLO}. It is the time-reversed diagram
of the photodissociation reaction considered in Ref.~\cite{Hammer:2011ye}.
At LO, the amplitude is
\begin{equation}
  \bar{\Gamma}^{i} = \frac{\boldsymbol{\epsilon^{i} \cdot p}}{M} \frac{\sqrt{Z_\sigma} e Q_c g_0 2 m_R}{\gamma_0^2
    + (\boldsymbol p - \frac{m}{M_{nc}} \boldsymbol k)^2} \ ,
\end{equation}
where $i$ is the photon polarization, $\boldsymbol p$ denotes the
relative momentum of the $nc$ pair and $\boldsymbol{k}$ the photon
momentum. Throughout this section we choose the $nc$ pair to be traveling
in $\hat{z}$ direction which means that $\boldsymbol{p} = |\boldsymbol{p}| \boldsymbol{e}_z$.  Since $m/M_{nc}$ is small and it follows from power counting that
$p \sim \gamma_0 \sim \Mlo$ and $k \sim \Mlo^2/\Mhi$, we can neglect the
recoil term $\sim \boldsymbol{p \cdot k}$ in the denominator.  By
averaging over the neutron spin and photon polarization and summing
over the outgoing $S$-wave spin we obtain at LO ($ \frac{m}{M_{nc}} k \ll p$):
\begin{align}
  \frac{d\sigma^{cap}}{d\Omega} &=
    \frac{m_R}{4\pi^2} \frac{k}{p} |\mathcal{M}^{(1/2)}|^2 \
    = \ \frac{e^2 Z_{eff}^2}{\pi m_R^2} \frac{p \gamma_0 \sin^2{\theta}}{(p^2+\gamma_0^2)} \ ,
\end{align}
with $k \approx (p^2+\gamma_0^2)/2 m_R$, $\bf \hat{k} \cdot \hat{p} = \cos\theta$, $Z_{eff} = (m_R/M) Q_c \approx 0.353$ and
\begin{align}
   |\mathcal{M}^{(1/2)}|^2 &=
   \frac{1}{2} \sum_{i,m_s,M} |\bar{\Gamma}^{i}|^2 \delta_{m_s,M} \ ,
\end{align}
where $m_s$ denotes the neutron spin and $M$ the $S$-wave
polarization. Since the neutron spin is unaffected by this reaction,
$m_s$ and $M$ have to be the same.  After integration over $d\Omega$
we get
\begin{align}
  \sigma^{cap} &=
    \frac{m_R}{\pi} \frac{k}{p} |\mathcal{M}^{(1/2)}|^2 \
    = \ \frac{8 e^2 Z_{eff}^2}{3 m_R^2} \frac{p \gamma_0}{(p^2+\gamma_0^2)}
    =  \ \frac{32 \pi \alpha Z_{eff}^2}{3 m_R^2} \frac{p \gamma_0}{(p^2+\gamma_0^2)} \ ,
\end{align}
with the fine-structure constant $\alpha = e^2/(4\pi)$.
Exploiting the detailed balance theorem, the capture
cross section $\sigma^{cap}$ can be related
to the photodissociation cross section
$\sigma^{dis}$~\cite{Baur:1986pd},
\begin{equation}
\label{eq:detailed_balance}
  \sigma^{cap} = \frac{2(2j_{^{17}\text{C}}+1)}{(2j_n+1)(2j_c+1)} \frac{k^2}{p^2} \ \sigma^{dis} \ = \ 2 \frac{k^2}{p^2} \ \sigma^{dis} \ .
\end{equation}
Our numerical results for the E1 capture into $^{17}$C
and photodissociation of
$^{17}$C obtained using Eq.~\eqref{eq:detailed_balance} at LO are shown in
Fig. \ref{fig:E1captureCrossSectionC17swave}.
At NLO, there is an additional contribution from the effective range $r_0$.
By assuming that $r_0$ scales as $1/\Mhi$, we can estimate the size of the
NLO contribution by multiplying the LO result by a factor of
$(1 \pm \gamma_0/M_{hi} )$ and add an error band to our LO results in
Fig.~\ref{fig:E1captureCrossSectionC17swave}.

\subsubsection{E1 capture into the $3/2^{+}$ and $5/2^{+}$ states}
\begin{figure}[t]
	\centering
		\includegraphics[width=0.5\textwidth]{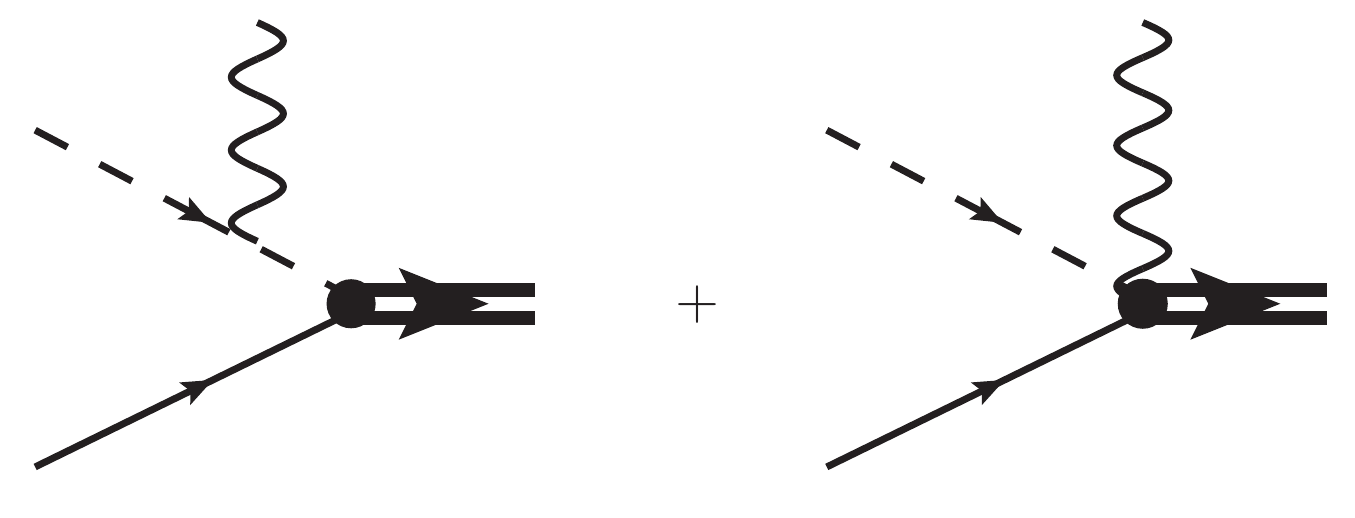}
	\caption{Relevant diagrams for E1 capture to $D$-wave states at LO. For a more detailed description of the lines, see Fig. \ref{fig:dimer-propagator}.}
	\label{fig:diagramsE1captureD}
\end{figure}

In this section, we calculate E1 neutron capture to the $3/2^+$
$D$-wave ground state and $5/2^+$ excited state of $^{17}$C. The
relevant diagrams that emerge from minimal substitution in our
Lagrangian~\eqref{eq:lagrangian_dwave} are shown in
Fig. \ref{fig:diagramsE1captureD} . They yield
\begin{align}
\notag
  \bar{\Gamma}^{i}_{m_s JM} &=  \sum_{m_{s'}
    m_l}
    \left(\left. \frac{1}{2} m_{s'} \ 2 m_l \right\vert
    J\, M \right) \sum_{\alpha \beta} \left(\left. 1
    \alpha \ 1 \beta \right\vert 2 m_l \right) \sqrt{Z_d} g_2 e Q_c \frac{2 m_R}{M} \times \\
    &\left[
  \frac{\left( \boldsymbol{p} - \frac{m}{M_{nc}} \boldsymbol{k} \right)_\alpha \left( \boldsymbol{p} - \frac{m}{M_{nc}} \boldsymbol{k} \right)_\beta }{\gamma_2^2 + \left(\boldsymbol{p} - \frac{m}{M_{nc}} \boldsymbol{k} \right)^2} \ \boldsymbol{\epsilon^{i} \cdot p} + \epsilon^{i}_\alpha \left(p_\beta - \frac{m}{2 M_{nc}} k_\beta \right) \right] \delta_{m_s m_s'}\ ,
\end{align}
with the charge of the core $Q_c$, the photon momentum
$\boldsymbol{k}$, the relative momentum of the incoming $nc$ pair
$\boldsymbol{p}$, the photon polarization $i$ and $JM$ denoting the spin
and polarization of the $D$-wave. Note that the neutron spin is unaffected by the E1 capture process up to this order. If we project out the $J=3/2$ part
of the amplitude $M^{(3/2)}$ and average (sum) over incoming (outgoing)
spins, respectively, we finally find the differential cross
section for the E1 capture process at LO ($ \frac{m}{M_{nc}} k \ll p$):
\begin{align}
 \frac{d\sigma^{cap}}{d\Omega} &= \frac{m_R}{4\pi^2} \frac{k}{p} \left| \mathcal{M}^{(3/2)}
  \right|^2 = \frac{15}{2 \pi} \frac{\left( p^2 + \gamma_2^2 \right)}{m_R^2 p}
  \frac{e^2 Z_{eff}^2}{-r_2 - \mathcal{P}_2 \gamma_2^2} X(\theta)
   = \frac{30 \alpha Z_{eff}^2}{-r_2 - \mathcal{P}_2 \gamma_2^2} \frac{\left( p^2 + \gamma_2^2 \right)}{m_R^2 p} X(\theta) \ ,
\end{align}
with the fine-structure constant $\alpha$, $Z_{eff} = (m_R/M) Q_c$,
\begin{align}
   &|\mathcal{M}^{(3/2)}|^2 =
     \frac{1}{2} \sum_{i,m_s,M} |\bar{\Gamma}^{i}_{m_s 3/2 M}|^2 \ ,
\end{align}
and
\begin{align}
  &X(\theta) =  \frac{1}{15} \left[ 2 p^2 (13 - \cos (2 \theta ))+ \frac{4 p^4 \sin ^2(\theta )}{\left(\gamma
   _2^2+p^2\right)} \left( \frac{p^2}{\left(\gamma
   _2^2+p^2\right)} + 2 \right) \right] \ .
\end{align}
After integrating over $d\Omega$ we find for the total cross section:
\begin{align}
 \sigma^{cap} = \frac{\alpha Z_{eff}^2}{-r_2 - \mathcal{P}_2 \gamma_2^2} \frac{32 \pi  p}{3 m_R^2} \frac{ \left(5 \gamma _2^4 + 11 p^4+ 14 \gamma _2^2 p^2\right)}{ \left(\gamma
   _2^2+p^2\right)}  \ .
\end{align}

From an experimental measurement of the capture (or dissociation)
cross section we can therefore extract the numerical value of the
combination of $D$-wave effective range parameters
$1/(-r_2 - \mathcal{P}_2 \gamma_2^2)$.  For the $5/2^+$ state we
project out the $J=5/2$ part of the amplitude $M^{(5/2)}$ and obtain:
\begin{align}
 \frac{d\sigma^{cap}}{d\Omega} &= \frac{m_R}{4\pi^2} \frac{k}{p} \left| \mathcal{M}^{(5/2)}
  \right|^2 = \frac{45}{4\pi} \frac{\left( p^2 + \gamma_{2'}^2 \right)}{m_R^2 p}
  \frac{e^2 Z_{eff}^2}{-r_{2'} - \mathcal{P}_{2'} \gamma_{2'}^2} X(\theta)
   = \frac{45 \alpha Z_{eff}^2}{-r_{2'} - \mathcal{P}_{2'} \gamma_{2'}^2} \frac{\left( p^2 + \gamma_{2'}^2 \right)}{m_R^2 p} X(\theta) \ ,
\end{align}
where $X(\theta)$ is the same as for the $J=3/2$ cross section.
After integrating over $d\Omega$ we find for the total cross section:
\begin{align}
 \sigma^{cap} = \frac{\alpha Z_{eff}^2}{-r_{2'} - \mathcal{P}_{2'} \gamma_{2'}^2} \frac{16 \pi  p}{m_R^2} \frac{ \left(5 \gamma _{2'}^4+ 11 p^4+ 14 \gamma _{2'}^2 p^2\right)}{ \left(\gamma
   _{2'}^2+p^2\right)} \ ,
\end{align}
which is the same result as the $J=3/2$ cross section multiplied by a
factor of $3/2$ and different numerical values for $\gamma_2$, $r_2$
and $\mathcal{P}_2$.

\subsection{M1 neutron capture on $^{16}$C}
\label{sec:m1-capture}
\begin{figure}[t]
  \centering \includegraphics[width=0.9\textwidth]{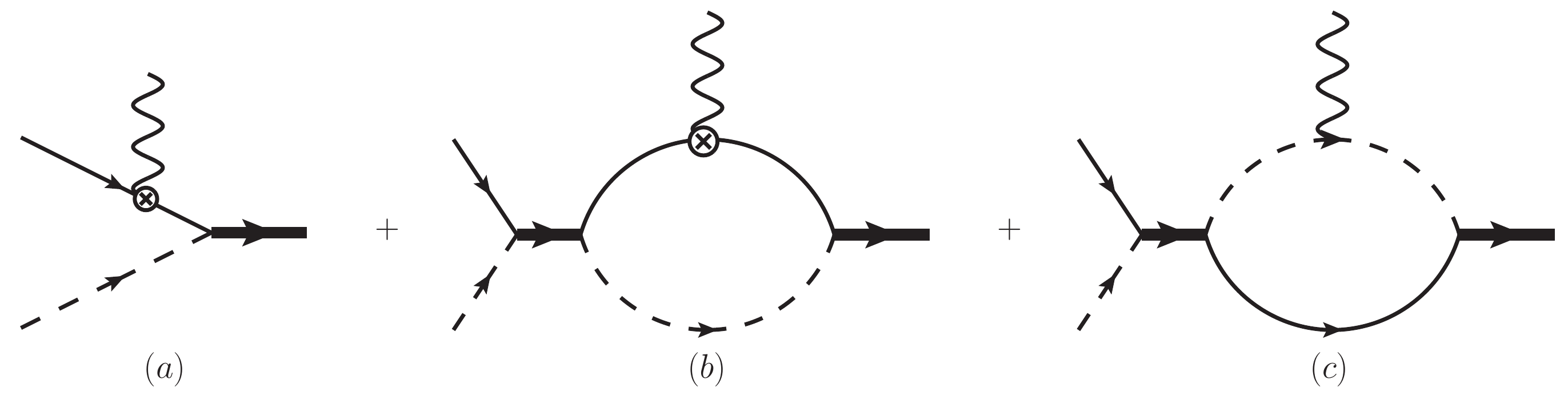}
  \caption{Relevant diagrams contributing to M1
    capture at LO. For a more detailed description of the lines, see
    Fig. \ref{fig:dimer-propagator} and
    \ref{fig:diagramsM1momentGammaV}.}
  \label{fig:diagramsM1capture}
\end{figure}

\subsubsection{M1 capture into the $1/2^{+}$ state}
Similar to E1 capture, we can calculate the M1 capture cross
section. The main difference between both processes is the parity
conservation in the M1 matrix element. Therefore, the loop diagram (b)
shown in Fig.~\ref{fig:diagramsM1capture} is also relevant at LO for
M1 capture since initial state interactions in the $S$-wave channel
have to be taken into account. Additionally, the photon now couples to
the magnetic moment of the halo neutron in diagrams
  (a) and (b).  In principle, we also need to consider diagrams which
  arise from minimal substitution.  This is shown in the third diagram
  (c) where the photon couples to the charged $^{16}$C core. In the
  $S$-wave case, however, diagram (c) yields no contribution to the M1
  capture process. For diagram (a) in
Fig.~\ref{fig:diagramsM1capture} we get:
\begin{align}
\label{eq:M1capA}
  \bar{\Gamma}^{(a)}_{i m_s m_{s'}} = -2 \sqrt{Z_\sigma} \kappa_n \mu_N g_0 m_R
  \frac{\sigma_j^{m_s m_{s'}} ({\bf k} \times {\boldsymbol \epsilon^i})_j}
  {\gamma_0^2
  + \left(\boldsymbol{p} - \frac{M}{M_{nc}} \boldsymbol{k} \right)^2} \ ,
\end{align}
with the Pauli matrices $\sigma_j$, the photon
polarization index $i$, and the relative momentum of the incoming $nc$ pair
$\boldsymbol{p}$.

Since the power counting stipulates $p \sim \gamma_0 \sim \Mlo$ and
$k \sim \Mlo^2/\Mhi$, we can neglect the
recoil term $\sim \boldsymbol{p}\cdot{\bf k}$ in the denominator
of Eq.~\eqref{eq:M1capA}.
\begin{align}
\label{eq:M1capALO}
  \bar{\Gamma}^{(a)}_{i m_s m_{s'}} =
  -2 \sqrt{2 \pi \gamma_0} \kappa_n \mu_N
  \frac{\sigma_j^{m_s m_{s'}} ({\bf k} \times {\boldsymbol \epsilon^i})_j}{\gamma_0^2 + p^2} \ .
\end{align}
Diagram (b) with the intermediate $S$-wave state yields
\begin{align}
  \bar{\Gamma}^{(b)}_{i m_s m_{s'}} =
  - \sqrt{Z_\sigma} g_0^3 \kappa_n \mu_N \frac{2 \pi}{g_0^2 m_R}
  \frac{\sigma_j^{m_s m_{s'}} ({\bf k} \times {\boldsymbol \epsilon^i})_j}{\frac{1}{a_0}
  -\frac{r_0}{2} p^2 + i p}
  \int{\frac{dl^3}{(2\pi)^3} \frac{2 m_R}{p^2 - l^2}
  \frac{2 m_R}{\gamma_0^2 + \left(\boldsymbol{l} + \frac{m_R}{m} \boldsymbol{k} \right)^2}} \ ,
\end{align}
with the loop momentum $\boldsymbol{l}$, which leads at LO to
\begin{align}
 \bar{\Gamma}^{(b)}_{i m_s m_{s'}} = 2 \sqrt{2 \pi \gamma_0} \kappa_n \mu_N \frac{\sigma_j^{m_s m_{s'}} ({\bf k} \times {\boldsymbol \epsilon^i})_j}{\gamma_0 + i p} \frac{1}{\gamma_0 - ip} = - \bar{\Gamma}^{(a)}_{i m_s m_{s'}} \ .
\end{align}
As a consequence, both diagrams cancel each other at LO. In coordinate
space, this process is given by an overlap integral between two
orthogonal wave functions. At NLO, there is an additional contribution from the effective range $r_0$ as discussed for the E1 capture process before,
which will give a correction of order $\gamma_0 r_0\approx 40\%$.
Moreover, a two-body current enters at NLO with an additional counter term
that has to be fixed from data, similar to the case of magnetic moments
discussed in Sec. \ref{sec:magetic-moment-swave}. This shows again that
counter terms play a more dominant role in the magnetic sector than in the
electric one.

\paragraph{Recoil corrections -}
Subleading recoil corrections are usually dropped in EFT
calculations for capture reactions such as this one. Taking recoil
corrections into account, the first diagram (a) will give non-zero
contributions to higher multipoles through higher partial waves in the
initial state. The second diagram (b) in Fig.~\ref{fig:diagramsM1capture}
contributes only when the core and the nucleon are in a relative
$S$-wave in the initial state.

The denominator in Eq.~\eqref{eq:M1capA} for diagram (a) can be
spherically expanded as
\begin{align}
\label{eq:M1capRecoil}
 \frac{1}{\gamma_0^2 + \left(\boldsymbol{p} - \frac{M}{M_{nc}}
  \boldsymbol{k} \right)^2} &=
                              - \sum_{l}{(2l+1) i^{2l} P_l(\hat{p} \hat{k}) \frac{M_{nc}}{2 M k p} \mathcal{R}e \bigg \{ Q_l\left( - \frac{M_{nc}}{2M pk} \left( p^2 +  \frac{M^2}{M_{nc}^2} k^2 + \gamma_0^2 \right) \right) \bigg \} } \ ,
\end{align}
where $Q_l(x)$ denotes the Legendre function of the second kind.

As an example, we consider the $S$-wave result for Eq. \eqref{eq:M1capRecoil}
\begin{align}
  - \frac{1}{a}
  \ln\left(1 - \frac{a}{\gamma_0^2 + \left(p + \frac{M}{M_{nc}} k \right)^2} \right) \ ,
\end{align}
with $a = M_{nc}/(4M kp)$, which is in perfect agreement with
Eq.~\eqref{eq:M1capALO} if we set $k \sim 0$ and expand the logarithm.

After averaging and summing over incoming and outgoing spins, respectively, we obtain for the differential cross section the general result:
\begin{align}
 \frac{d\sigma^{cap}}{d \Omega} = \frac{m_R}{4 \pi^2} \frac{k}{p} |\mathcal{M}^{(1/2)}|^2 =
    \frac{m_R}{m^2}
    \frac{k^3}{p}
     \frac{4 \alpha\kappa_n^2 \gamma_0}
    {\left[\gamma_0^2 + \left(\boldsymbol{p} - \frac{M}{M_{nc}}
    \boldsymbol{k} \right)^2 \right]^2}\ ,
\end{align}
with the fine structure constant $\alpha$ and 
\begin{align}
   |\mathcal{M}^{(1/2)}|^2 &=
   \frac{1}{2} \sum_{i,m_s,m_{s'}} |\bar{\Gamma}^{(a)}_{i m_s m_{s'}}|^2 \ .
\end{align}

\subsubsection{M1 capture into the $3/2^{+}$ and $5/2^{+}$ states}
\begin{figure}[t]
  \centering \includegraphics[width=0.65\textwidth]{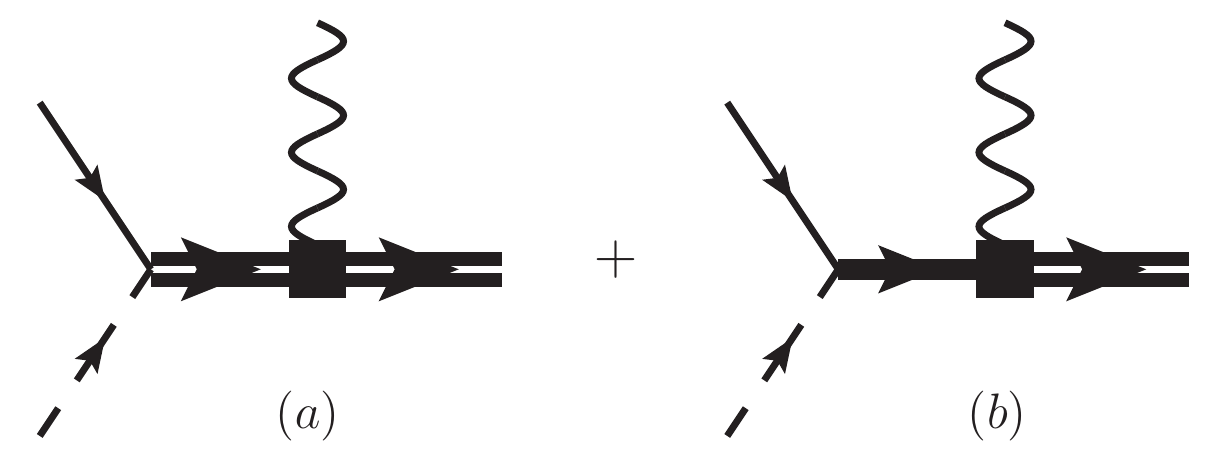}
  \caption{Relevant diagrams contributing to M1
      capture into $D$-wave at up to NLO. The thick double line
      denotes the dressed $D$-wave dimer and the thick single line the
      dressed $S$-wave dimer. For a description of the other
      lines, see Figs. \ref{fig:dimer-propagator} and
      \ref{fig:diagramsM1momentGammaV}. The solid squares denote
      different vertices from two-body currents.}
  \label{fig:diagramsM1captureD}
\end{figure}
In this section, we calculate M1 neutron capture
from the continuum into the $3/2^+$ $D$-wave ground state or $5/2^+$
excited state of $^{17}$C. Compared to the $1/2^+$ case in the
previous section, there are additional contributions from two-body
currents for the $D$-wave case at LO and NLO:
\begin{align}
\notag
  \mathcal{L}_{M} =&  - \mu_N L^{d'd}_{M1cap} d^\dagger_{m'} d'_{m} B_i \left(\frac{5}{2} m 1 i \left| \frac{3}{2} m' \right.\right)  - \mu_N L^{dd}_{M1cap} d^\dagger_{m'} d_{m} B_i \left(\frac{3}{2} m 1 i \left| \frac{3}{2} m' \right.\right)\\
  \label{eq:M1capD}
  &  - \mu_N L^{d'd'}_{M1cap} d'^\dagger_{m'} d'_{m} B_i \left(\frac{5}{2} m 1 i \left| \frac{5}{2} m' \right.\right) -  \mu_N L^{\sigma d}_{M1cap} d^\dagger_{m'} \sigma_{m} B_i \left(\frac{1}{2} m 1 i \left| \frac{3}{2} m' \right.\right)~.
\end{align}
By rescaling the fields to absorb unnaturally large coupling
constants, leading to $[\tilde{\sigma}] = 2$, $[\tilde{d}] = [\tilde{d'}] = 0$,
and using naive dimensional analysis for the rescaled fields
\cite{Beane:2000fi}, we find 
$L_{M1cap}^{d'd} \sim \Mhi^3 l_{M1cap}^{d'd} g_{2'} g_2 m_R^2$,
$L_{M1cap}^{d^{(\prime)} d^{(\prime)}} \sim \Mhi^3 l_{M1cap}^{d^{(\prime)} d^{(\prime)}} g^2_{2^{(\prime)}} m_R^2$ and
$L_{M1cap}^{\sigma d} \sim \Mhi l_{M1cap}^{\sigma d} g_0 g_2 m_R^2$
with the constants $l_{M1cap}^{\cdots}$ all of order one.
The corresponding diagrams are shown in
Fig. \ref{fig:diagramsM1captureD}. The first diagram (a) represents
the first three terms in Eq. \eqref{eq:M1capD} where the two-body
current is between two $D$-wave states. This is the LO contribution to
the M1 capture process. The second diagram (b) belongs to the last
term in Eq. \eqref{eq:M1capD} and is only relevant for the $3/2^+$
ground state. This yields an NLO contribution. The diagram (a) in
Fig. \ref{fig:diagramsM1capture}, where the photon couples to the
magnetic moment of the neutron, contributes at N$^2$LO and the two
loop diagrams at N$^3$LO. Since we get additional counter terms
$L_{M1cap}$ that have to be matched to data, predictions for the M1
capture in $D$-wave case become even more complicated.  For that
reason, we concentrate on the LO result which yields for the $5/2^+$
excited state:
\begin{align}
\notag
  \bar{\Gamma}_{m_s \frac{5}{2} M}^{i} = \sum_{m_s
    m_l}
    \left(\left. \frac{1}{2} m_s \ 2 m_l \right\vert
    \frac{5}{2} M' \right) \sum_{\alpha \beta} \left(\left. 1
    \alpha \ 1 \beta \right\vert 2 m_l \right) \frac{{p}_\alpha {p}_\beta}{\sqrt{r_{2'} + \mathcal{P}_{2'} \gamma_{2'}^2}} \times \\
   \mu_N \sum_{M' \gamma} \left(\frac{5}{2} M' 1 \gamma \left| \frac{5}{2} M \right.\right) ({\bf k} \times {\boldsymbol \epsilon^i})_\gamma \
   \frac{  \tilde{L}_{M1cap}^{d'd'}}{\frac{1}{a_{2'}} - \frac{r_{2'}}{2} p^2 + \frac{\mathcal{P}_{2'}}{4} p^4 } \ ,
\end{align}
with the $D$-wave polarizations $\alpha$ and $\beta$, the photon
momentum $\boldsymbol{k}$, photon polarization $i$, the relative
momentum of the incoming $nc$ pair $\boldsymbol{p}$, and we have defined
$\tilde{L}_{M1cap}^{d'd'} = \frac{(15 \pi)^{3/2}}{m_R^2
  g_{2'}^2}L_{M1cap}^{d'd'}$.
For the $3/2^+$ ground state we obtain:
\begin{align}
\notag
  \bar{\Gamma}_{m_s \frac{3}{2} M}^{i} = \left(\left. 1
    \alpha \ 1 \beta \right\vert 2 m_l \right) \frac{{p}_\alpha {p}_\beta \mu_N  ({\bf k} \times {\boldsymbol \epsilon^i})_\gamma}{{\sqrt{r_{2} + \mathcal{P}_{2} \gamma_{2}^2}}} \left[\left(\left. \frac{1}{2} m_s \ 2 m_l \right\vert
    \frac{3}{2} M' \right) \left(\frac{3}{2} M' 1 \gamma \left| \frac{3}{2} M \right.\right) \frac{  \tilde{L}_{M1cap}^{dd}}{\frac{1}{a_{2}} - \frac{r_{2}}{2} p^2 + \frac{\mathcal{P}_{2}}{4} p^4 } \right. \\
    +  \left. \left(\left. \frac{1}{2} m_s \ 2 m_l \right\vert
    \frac{5}{2} M' \right) \left(\frac{5}{2} M' 1 \gamma \left| \frac{3}{2} M \right.\right) \frac{  \tilde{L}_{M1cap}^{d'd}}{\frac{1}{a_{2'}} - \frac{r_{2'}}{2} p^2 + \frac{\mathcal{P}_{2'}}{4} p^4 } \right] \ ,
\end{align}
where we have implicitly summed over repeated indices and we have defined $\tilde{L}_{M1cap}^{dd} = \frac{(15 \pi)^{3/2}}{m_R^2 g_{2}^2}L_{M1cap}^{dd}$ and $\tilde{L}_{M1cap}^{d'd} = \frac{(15 \pi)^{3/2}}{m_R^2 g_{2'} g_2}L_{M1cap}^{d'd}$.
The differential cross section for the M1 capture process at LO for $J=3/2$ or $5/2$ is then given by:
\begin{align}
 \frac{d\sigma^{cap}}{d \Omega} = \frac{m_R}{4 \pi^2} \frac{k}{p} |\mathcal{M}^{(J)}|^2 ~, \quad \text{with} \quad |\mathcal{M}^{(J)}|^2 &=
   \frac{1}{2} \sum_{i,m_s,M} |\bar{\Gamma}_{m_s J M}^{i}|^2 \ .
\end{align}
Since we need at least four additional input parameters to make predictions for the
M1 capture process into the $D$-wave state already at LO, numerical predictions are currently not possible.
This shows the limitations of Halo
EFT for higher partial waves especially in the magnetic sector.

\section{Summary}
\label{sec:summary}
Halo nuclei are weakly bound systems of a tightly bound core nucleus
and a small number of valence nucleons. Their structure can be probed
experimentally by measuring capture reactions, dissociation cross
sections, and charge radii.  In this work, we have discussed these
observables for $S$- and $D$-wave halo states using the framework of
Halo EFT.

We have considered the nucleus $^{17}$C as a halo nucleus consisting
of a $^{16}$C core and a neutron.  $^{17}$C is an interesting halo
candidate since it has three $S$- and $D$-wave neutron-core states
with small neutron separation energies in its spectrum. We have
calculated the key observables relevant to this system, including
radii, magnetic moments as well as electric and magnetic transition
rates.  Moreover, we showed that capture reactions can provide insight
into the continuum properties of the neutron-$^{16}$C system.

We found that predictions of many observables for states with angular
momentum larger than zero need additional input parameters, beyond the
neutron separation energy. This limits the predictive power of Halo
EFT for such states.  However, these counterterms can be matched to
experiment or other theoretical calculations. For example, the
counterterms appearing in the expressions for the $S$- to $D$-wave
transitions can be determined in this way. Coupled-cluster calculations
for $^{17}$C were carried out in Ref.~\cite{Kanungo:2016tmz} using
effective interactions derived from first principles, and this
approach could be extended to calculate the transitions in our work.
The results could then be used to predict capture cross sections since
the counterterms in capture cross sections and transition strengths
are related. This strategy would provide insights into the continuum
properties of ${}^{17}$C based on a combination of halo EFT and the
shell model.  Alternatively, one can eliminate unknown counterterms by
considering correlations between different observables. These
correlations can be used to test the consistency between different ab
initio calculations and/or experimental data. The structure of such
correlations is universal in the sense that it is independent of the
specific neutron separation energies and applies to all states with
the same quantum numbers.  As a consequence, Halo EFT is complementary
to ab initio approaches by exploiting universal correlations driven by
the weak binding.

Some of the observables discussed in this work have been studied
extensively in the case of the deuteron which can be considered the
lightest halo nucleus, consisting of a neutron and a proton
core~\cite{Chen:1999tn,Chen:1999vd}.
One-neutron halo nuclei can therefore have similar electromagnetic
properties to the deuteron. For example, the expression for the LO
charge radius of an $S$-wave neutron halo nucleus shown in
Eq.~\eqref{eq:rE-swave} is the same as for the deuteron. However, the
deuteron consists of two spin-1/2 particles and interacts resonantly
in the spin-triplet and spin-singlet $S$-wave channels. This leads to
a relatively large M1 capture cross section between the unbound
spin-singlet and the spin-triplet channel in which the deuteron
resides.  The absence of a second resonantly interacting channel leads
a strong suppression of magnetic capture in the case of ${}^{17}$C.

We hope that our investigation will motivate further theoretical and
experimental investigations of $^{17}$C.  The expressions presented in
this paper should be useful for the analysis of experimental and/or ab
initio data on ${}^{17}$C in order to establish the halo nature of
$^{17}$C. The combination of Halo EFT and ab initio calculations as was
done in Refs. \cite{Hagen:2013jqa,Ryberg:2014exa,Zhang:2014zsa} could
provide insights into the continuum properties of ${}^{17}$C and
should facilitate a test of the power counting that was used in this
work.

Future extensions of our calculation to NLO and beyond
would improve this comparison quantitatively, but a growing number of
counterterms may invalidate this advantage.

\acknowledgements We acknowledge useful discussions with Thomas
Papenbrock and Wael Elkamhawy.  JB thanks the University of Tennessee,
Knoxville and the Joint Institute for Nuclear Physics and Applications
for their hospitality and partial support.  This work has been
supported by Deutsche Forschungsgemeinschaft under grant SFB 1245, by
the BMBF under grant No. 05P15RDFN1, by the Office of Nuclear Physics,
U.S.~Department of Energy under Contract No. DE-AC05-00OR22725 and the
National Science Foundation under Grant No. PHY-1555030.


%

\end{document}